\documentclass{llncs}
\usepackage{makeidx}  
\usepackage{cite} 
\usepackage{hyperref}
\usepackage[utf8]{inputenc} 
\usepackage[misc]{ifsym}
\usepackage{tikz}
\usetikzlibrary{arrows,shapes,positioning,shadows,trees,calc,decorations.markings}
\begin{document}
\frontmatter          

\mainmatter              
\title{Brain Tumor Segmentation and Survival Prediction using 3D Attention UNet} 
\titlerunning{}  
%
\author{Mobarakol Islam\inst{1,2}$^{\textrm{(\Letter)}}$ \and Vibashan VS\inst{2,3} \and V Jeya Maria Jose \inst{2,3} \and Navodini Wijethilake\inst{2,4} \and Uppal Utkarsh\inst{2,5} \and 
Hongliang Ren \inst{2}$^{\textrm{(\Letter)}}$}
\authorrunning{Mobarakol et al.} 
%
\tocauthor{Example Author} 
%
\institute{NUS Graduate School for Integrative Sciences and Engineering, NUS, Singapore
\and
Dept. of Biomedical Engineering, National University of Singapore, Singapore\\
\and
Dept. of Instrumentation and Control Engineering, NIT, Tiruchirappalli, India\\
\and
Dept. of Electronics and Telecommunications, University of Moratuwa, Srilanka\\
\and
Dept. of Electrical Engineering, Punjab Engineering College, Chandigarh, India\\
\email{mobarakol@u.nus.edu, ren@nus.edu.sg }}

\maketitle              

\begin{abstract}
In this work, we develop an attention convolutional neural network (CNN) to segment brain tumors from Magnetic Resonance Images (MRI). Further, we predict the survival rate using various machine learning methods. We adopt a 3D UNet architecture and integrate channel and spatial attention with the decoder network to perform segmentation. For survival prediction, we extract some novel radiomic features based on geometry, location, the shape of the segmented tumor and combine them with clinical information to estimate the survival duration for each patient. We also perform extensive experiments to show the effect of each feature for overall survival (OS) prediction. The experimental results infer that radiomic features such as histogram, location, and shape of the necrosis region and clinical features like age are the most critical parameters to estimate the OS.

\end{abstract}
\section{Introduction} 
Gliomas develop from glial cells, are the most common brain tumor with the highest mortality rate. The mean occurrence of gliomas is close to 190,000 cases annually in worldwide \cite{castells2009automated}. The average survival time of the glioma patients remains at approximately 12 months \cite{furnari2007malignant}, and nearly 90\% of patients are dead after 24 months of surgical resection \cite{louis20072007}. Early detection, automatic delineation, and volume estimation are vital tasks for survival prediction and treatment planning. However, gliomas are often difficult to localize and delineate with conventional manual segmentation due to their high variation of shape, location, and appearance. In addition, close supervision from a human expert is required to manually annotate the segmentation of tumor tissue, which is time-consuming and tedious. Automatic segmentation and survival rate prediction models will help the diagnosis and treatment to be much accurate and faster.

In recent years, deep learning has dominated most of the tasks like segmentation \cite{islam2018glioma, islam2018ichnet, islam2018ischemic}, tracking \cite{islam2019learning, islam2019real}, and classification \cite{li2014medical} in medical image analysis. Many studies are for brain tumor segmentation, and survival prediction utilizes deep learning techniques, especially convolutional neural network (CNN). In this paper, we design a 3D attention based UNet \cite{ronneberger2015u} for brain tumor segmentation from MR images. To predict the survival days for each patient, we extract shape and geometrical features and combine them with clinical features and train to analyze the performance of various regression techniques like Support Vector Machine (SVM), Artificial Neural Network (ANN), Random Forest and XGBoost.

\section{Methods} 

\subsection{Segmentation}
\begin{figure}[!htbp]
  \centering
  \includegraphics[width=1\linewidth]{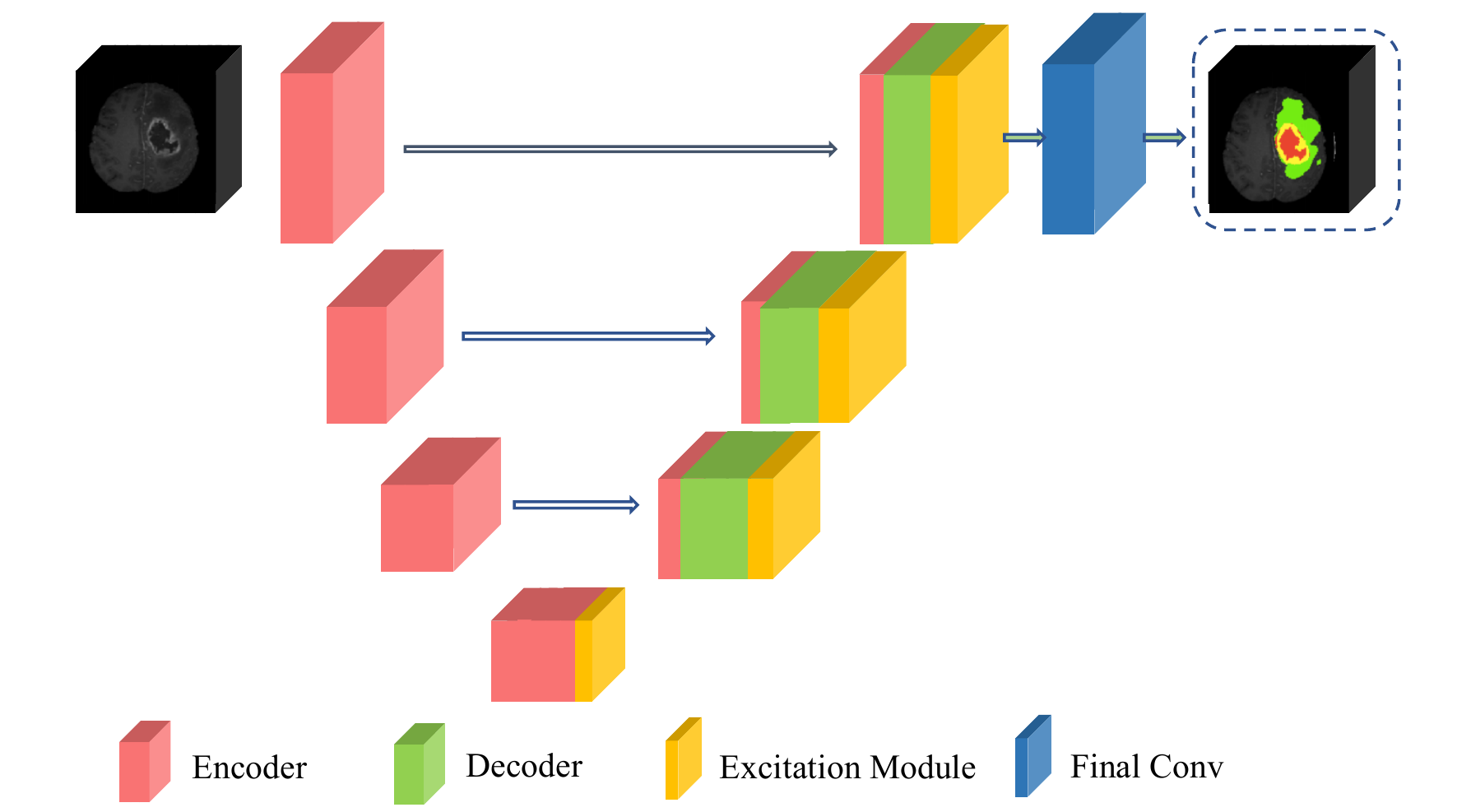}
  \caption{ Our proposed segmentation architecture 3D attention UNet by composing of sequential channel and spatial attention mechanism.}
  \label{fig:3d_atten1}  
\end{figure}

\begin{figure}[!htbp]
  \centering
  \includegraphics[width=1\linewidth]{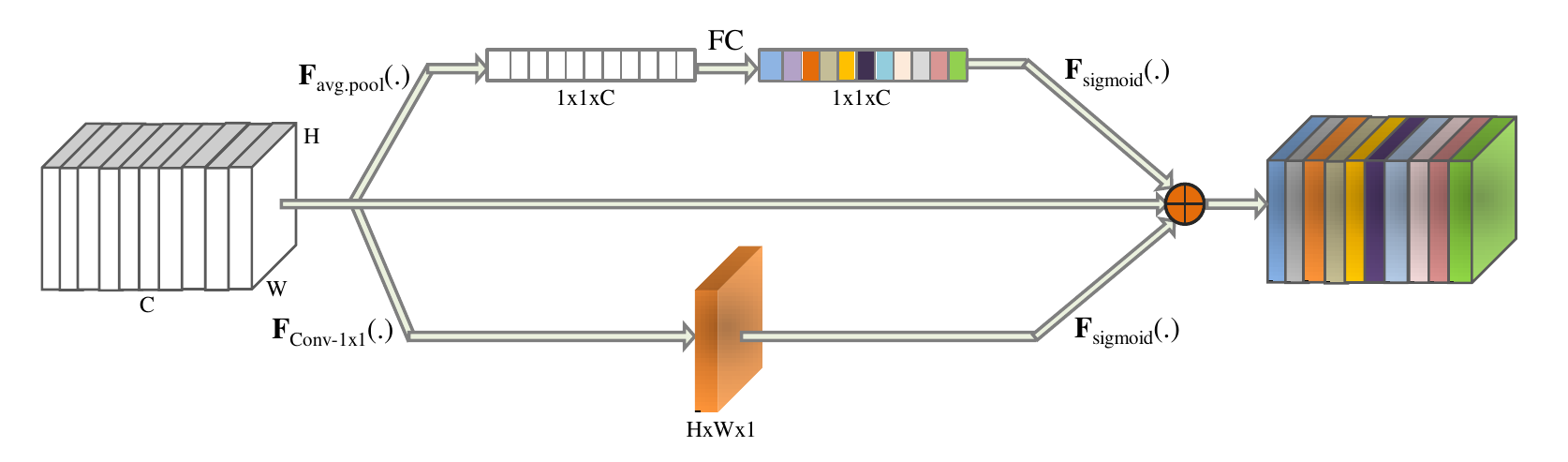}
  \caption{Visual representation of the 3D spatial and channel attention with skip connection.}
  \label{fig:3d_atten}  
\end{figure}

We adopt the UNet \cite{ronneberger2015u} architecture and convert it to 3D and integrate the 3D attention module with the decoder blocks. Further, we propose a 3D attention model with decoder blocks to enhance segmentation prediction\cite{hu2018squeeze}. Our proposed attention module consists of a channel and spatial attention in parallel with skip connection.   Nonetheless, fusing parallelly exciting features may create inconsistency in feature learning. Integrating skip connection reduces this redundancy and sparsity of the network, as illustrated in Fig \ref{fig:3d_atten}. The overall architecture is illustrated in Fig \ref{fig:3d_atten1}.

\subsubsection{3D Skip Attention Unit}
Spatial and channel attention enhances the quality of encoding throughout its feature hierarchy. Therefore we introduce 3D attention units to generate 3D spatial and channel attention by exploiting 3D inter-spatial and inter-channel feature relationships (as illustrated in Fig. \ref{fig:3d_atten}). To obtain the 3D attention map, we first perform a 1x1xC convolution to aggregate all spatial feature correlations into the HxWx1 dimension. In parallel, we perform average pooling and feed it to the neural network to get the 1x1xC channel correlation. The encoded 3D attention map encodes rich spatial and channel attention. Further, we fuse skip-connection to reduce sparsity and singularity caused by these parallel excitations. Moreover, integrating skip connection makes the learning more generic and enhancing the segmentation prediction.

\subsection{Survival Prediction}

\subsubsection{Feature Extraction}
Features that give information about the geometry, fractal nature of the tumor hold an important role in the number of days of survival as in our previous work \cite{islam2018glioma}. The combination of features used in \cite{islam2018glioma} produces the best accuracy for BraTS 2018 overall survival (OS) prediction task for the validation task. However, due to over-fitting of the data on the regression model, the method failed during the BraTS 2019 test phase. Therefore, the same combination of features is used in this work improvising the learning methods. The first axis, second axis, and third axis coordinates and lengths are extracted as geometrical features. In addition, centroid coordinates, eigenvalues, meridional and equatorial eccentricity, fractal dimensions, histogram features of the image including entropy, skewness, and kurtosis are also extracted for necrosis, tumor core, and whole tumor. All the features are normalized to 0-1 range to avoid the magnitude differences.

\subsubsection{Feature Selection }
To optimize the regression model prediction, we need to feed the model with the most decisive features for survival prediction. Thus, we explore recursive feature elimination (RFE) for feature ranking. The core idea of this method is to obtain the most significant features. The number of features is increased one by one to find the optimum number of features, which involves mostly for the overall survival (OS) prediction task. 

\subsubsection{Regression Model}

We utilize the state-of-art XGBoost regression model \cite{chen2016xgboost} on the selected features, to predict the overall survival (OS).  We tune the hyperparameters such as maximum tree depth, learning rate, the degree of verbosity, L1, and L2 regularization terms on weights to obtain the best performing model.  As L1 and L2 terms control the sparsity and over-fitting, the utilization of regularization terms is an advantage in regression tasks. We also apply several other machine learning tools that are used commonly for regression tasks. For example, multi-layer perceptron (MLP), support vector machine (SVM) \cite{suykens1999least} and random forest (RF) \cite{liaw2002classification}.

\section{Experiments}
\subsection{Dataset}
Brain tumor dataset of BraTS 2019 \cite{menze2015multimodal, bakas2017advancing, bakas2017segmentation,bakas2017segmentationlgg, bakas2018identifying} is used to conduct all the experiments in work. The train set of BraTS 2019 consists of 335 cases with high and low-grade glioma of 259 and 76, respectively. There are 125 and 166 cases in the validation and test set, respectively. Each case contains MRI images of 4 modalities - a) native (T1) b) post-contrast T1-weighted (T1Gd), c) T2-weighted (T2), and d) T2 Fluid Attenuated Inversion Recovery (T2-FLAIR). The voxel size of the modality is 240x240x155. There is also a segmentation annotation in the train set where 3 regions are labels as 1, 3, and 4 pixels values. The annotated labels denote the necrotic and non-enhancing tumor core (NCR/NET: 1), the peritumoral edema (ED: 2), and GD-enhancing tumor (ET: 4).

\subsection{Implementation details}
Our model is trained using Pytorch \cite{paszke2017automatic} deep learning framework. The learning rate and weight decay are adopted as 0.00015 and 0.005, respectively. We use the ADAM optimizer to train the model. Two NVIDIA GTX 1080 Ti 12 GB GPUs are exploited to conduct all the experiments in this work.

As a model input, we use the 3D voxel in 4 available modalities by cropping the brain region. The study \cite{myronenko20183d} utilizes the 3D data of 128x128x128 to fit the GPU memory and achieves the best accuracy in BraTS 2018 challenge. We apply a random crop of 128x192x192 and mean normalization inside the data loader to prepare our model input.

\section{Results}

\begin{figure}[!h]
  \centering
  \includegraphics[width=0.95\linewidth,trim={4cm 0 5cm 0}]{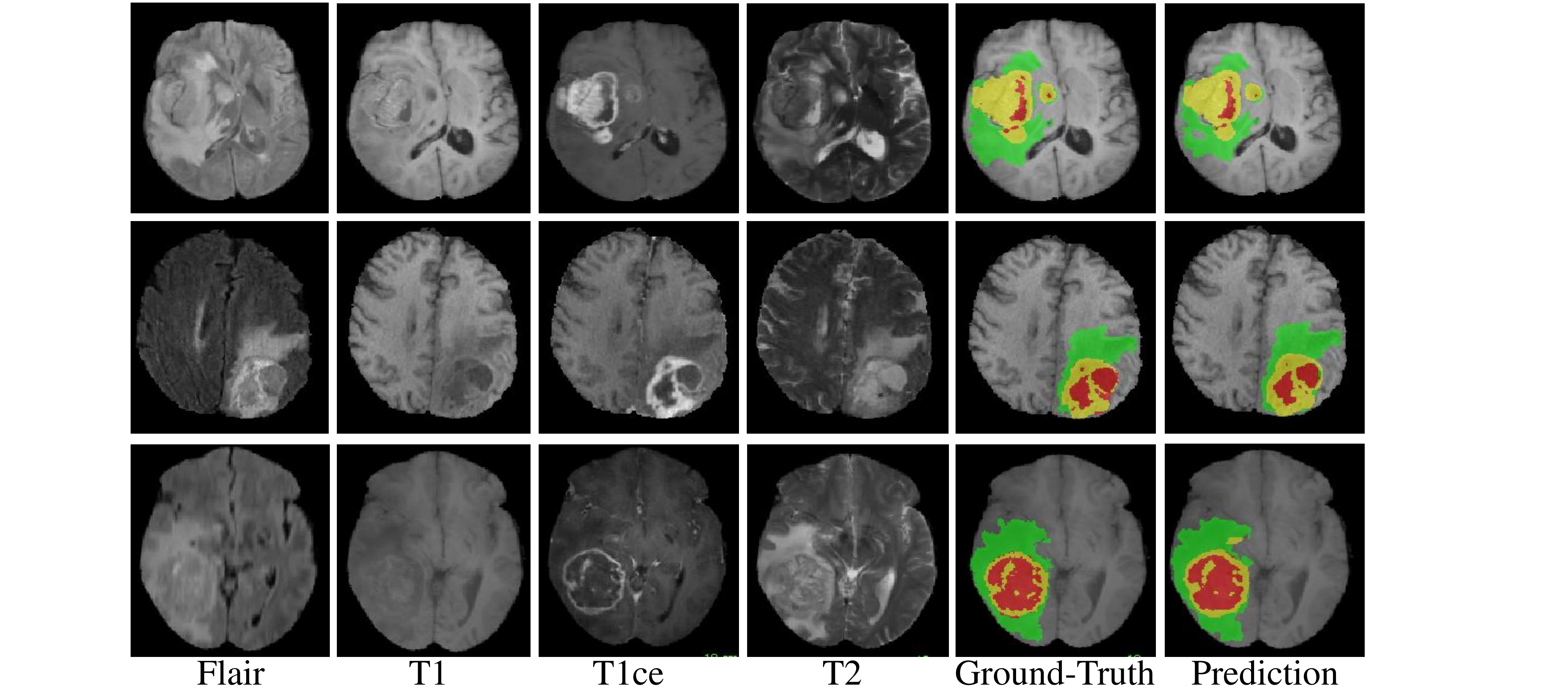}
  \caption{Flair, T1, T1ce, and T2 modalities of the brain tumor visualized with the Ground-Truth and Predicted segmentation of tumor sub-regions for BraTS 2019 cross-validation dataset. Red label: Necrosis, yellow label: Edema and Green label: Edema.}
  \label{fig:valid_visual}  
\end{figure}

\begin{figure}[!h]
  \centering
  \includegraphics[width=0.4\linewidth,trim={4cm 0 5cm 0}]{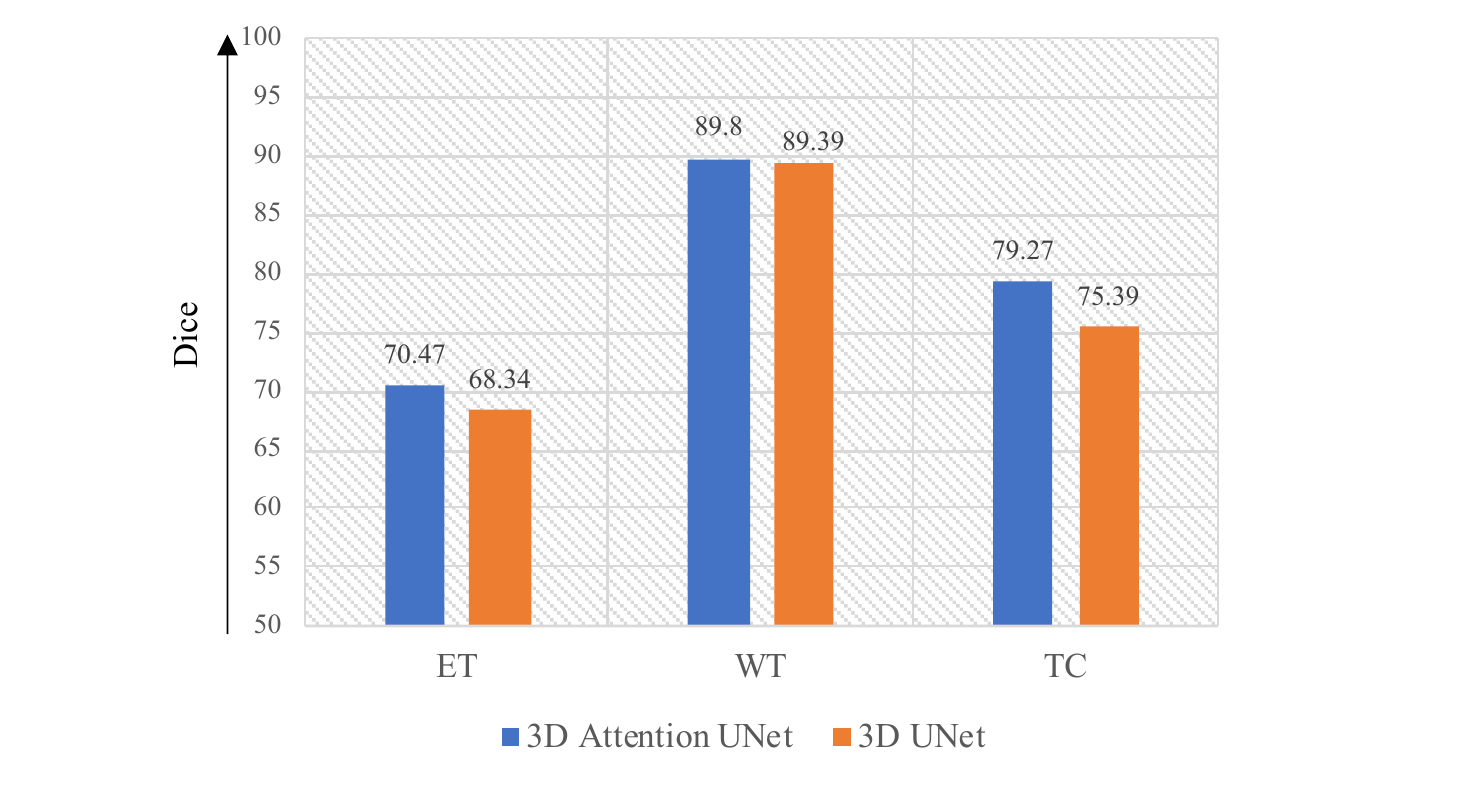}
  \caption{Performance comparison between our proposed model 3D attention UNet and original model 3D UNet.}
  \label{fig:result_comp}  
\end{figure}

To evaluate our model prediction, we submit the model prediction into the BraTS 2019 portal and obtain several measurement metrics such as Dice, Hausdorff, Sensitivity, and Specificity. The performances of the BraTS 2019 validation set are demonstrated in Table \ref{table:valid}. The visualization of the validation set prediction is illustrated in Fig. \ref{fig:valid_visual}. The performance graph of our proposed 3D attention UNet and original 3D UNet is plotted in Fig. \ref{fig:result_comp}. It is clearly shown that 3D attention UNet outperforms the original model for all the regions such as ET, WT, and TC. 

\begin{figure}[!h]
  \centering
  \includegraphics[width=.9\linewidth]{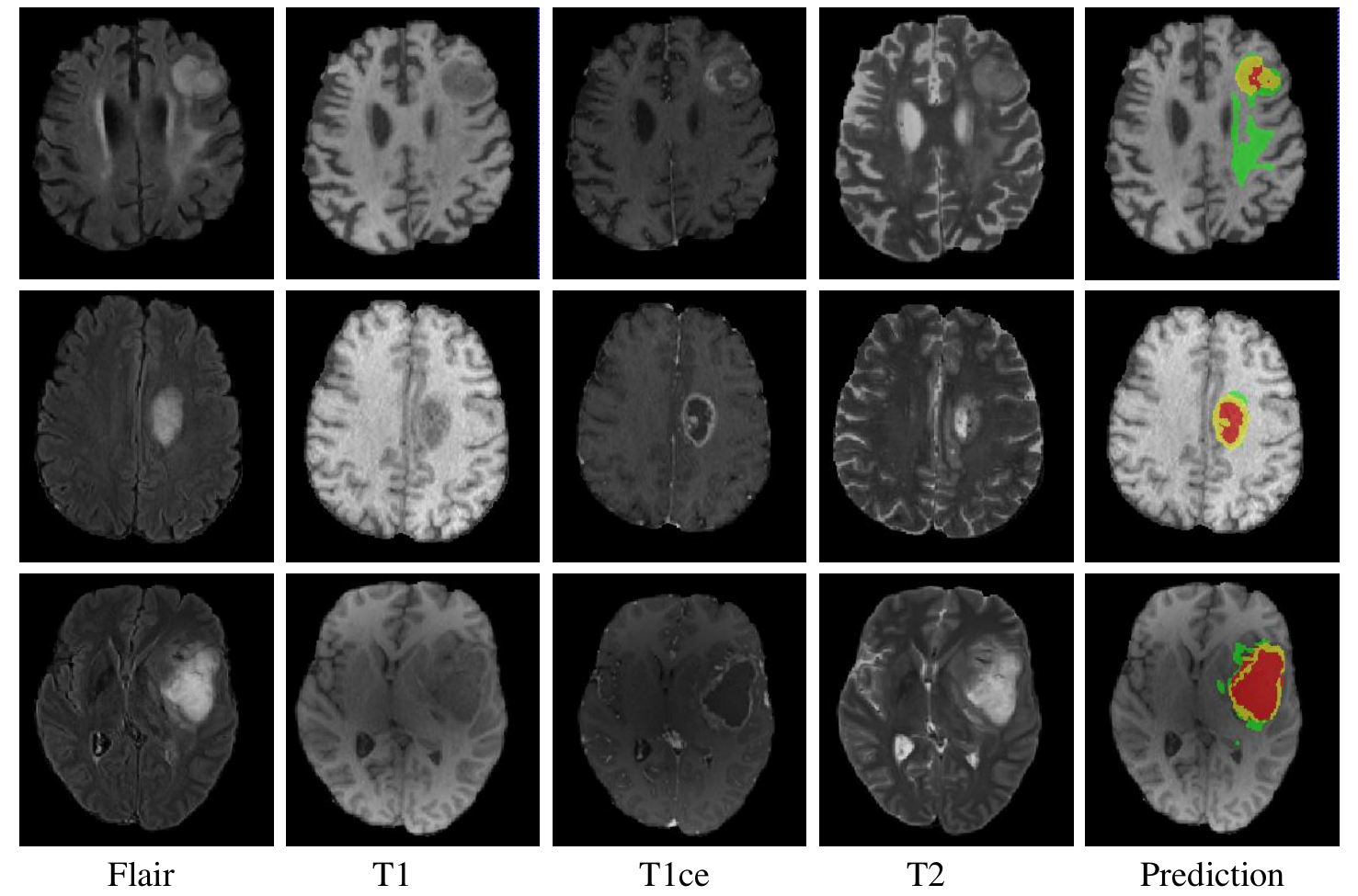}
  \caption{Flair, T1, T1ce and T2 modalities of the brain tumor visualized with the Ground-Truth and Predicted segmentation of tumor sub-regions for BraTS 2019 testing dataset. The annotation color can be interpreted as red - necrosis, yellow - enhance tumor, and green - edema.}
  \label{fig:test_vis}  
\end{figure}

The quantitative results for the BraTS 2019 test set are showed in Table \ref{table:test}. In Fig. \ref{fig:test_vis}, we can infer the prediction of our model for the BraTS 2019 testing dataset.

\begin{table}[!h]
 \caption{Dice, Hausdorff, Sensitivity, and Specificity metrics evaluation of BraTS 2019 validation set for segmentation task.}
\begin{center}
\label{table:valid}
\begin{tabular}{c|c|c|c|c|c|c|c|c|c|c|c|c}
\cline{1-13}
 & \multicolumn{3}{|c|}{Dice} & \multicolumn{3}{|c}{Hausdorff} & \multicolumn{3}{|c}{Sensitivity}& \multicolumn{3}{|c}{Specificity}\\
\cline{1-13}
{}      &ET     &WT     &TC     &ET &WT &TC &ET &WT &TC &ET &WT &TC\\ \hline
Mean    &0.704    &\textbf{0.898}  &0.792  &7.05 &6.29  &8.76  &0.751 &0.900 &0.816 &0.998 &0.994 &0.996 \\ \hline
StdDev    &0.311    &0.070  &0.190 &13.09 &11.14 &13.95 &0.284 &0.086 &0.191  &0.003    &0.005    &0.007 \\ \hline
Median    &0.835    &0.917  &0.868 &2.23  &3.31  &4.24  &0.859 &0.926 &0.894 &0.999    &0.996    &0.998 \\ \hline
\end{tabular}
\end{center}
\end{table}

\begin{table}[!h]
 \caption{Dice and Hausdorff metrics evaluation of BraTS 2019 testing set for segmentation task.}
\begin{center}
\label{table:test}
\begin{tabular}{c|c|c|c|c|c|c}
\cline{1-7}
 & \multicolumn{3}{|c|}{Dice} & \multicolumn{3}{|c}{Hausdorff}\\
\cline{1-7}
{}      &ET     &WT     &TC     &ET     &WC     &TC \\ \hline
Mean    &0.7780    &\textbf{0.8689 }   &0.7771    &3.6730    &7.3071    &6.8196    \\ \hline
StdDev    &0.2111    &0.1496    &0.2873    &6.1930    &13.6302    &11.3926    \\ \hline
Median    &0.8389    &0.9130    &0.8949    &2.0000    &3.6055    &3.08114    \\ \hline
\end{tabular}
\end{center}
\end{table}

\subsection{Survival Prediction}
 \begin{table}[!htbp]
\caption{Performance comparison of SVM, XGBoost, MLP and Random Forest (RF) on validation set for overall survival prediction. MSE and stdSE denotes as the mean square error and standard deviation of the predicted survival days.}
\centering
\label{table:training_comp}
\begin{tabular}{c|c|c|c|c}
\hline
Method     &Accuracy    &MSE    &MedianSE    &stdSE        \\ \hline
 XGBoost\cite{chen2016xgboost}&\textbf{42.86\%}    &{110012.835}    &{38444.333} &207273.871     \\ \hline
 MLP \cite{ruck1990multilayer}      &41.4\%    &\textbf{102839.036}    &49823    &138563.601    \\ \hline
Random Forest \cite{liaw2002classification}  &35.6\%    &268310.586    &58369.883    &12603.182        \\ \hline
SVM \cite{suykens1999least}   &32.9\% &107569.325    &72686.271    &106573.219        \\ \hline
\end{tabular}
\end{table}

\begin{table}[!h]
\caption{Quantitative results for XGBoost based survival prediction on the BraTS19 validation and test dataset}
\centering
\label{table:results_valid_test}
\begin{tabular}{c|c|c|c|c|c|c}
\hline
Dataset &Cases     &Accuracy    &MSE    &MedianSE    &stdSE    &SpearmanR    \\ \hline
Valid & 94 &\textbf{48.3\%}    &{127478.649}    &{35101.147} &211645.67  &0.187    \\ \hline
Test & 107    & 38.3\%    & 417633.26    & 68150.079    & 1215799.813&    0.238    \\ \hline
\end{tabular}
\end{table}

 Several state-of-the-art regression models are used to estimate the survival rate in our study. There are 125 cases in the validation set, but only 29 anonymous cases are chosen to validate the model in BraTS 2019 evaluation portal. We have done 4-fold cross-validation to evaluate the regression model on the training dataset. Table \ref{table:training_comp} shows the performance comparison among all the models. We observe that XGBoost outperforms all other regression models with the highest accuracy where MLP achieves the lowest MSE. We select XGBoost to evaluate the validation and test set by considering performance. Table \ref{table:results_valid_test} shows the XGBoost OS performance on BraTS 2019 validation and test dataset.

\section{Discussion}

\begin{figure}[!htbp]
  \centering
  \includegraphics[width=1\linewidth]{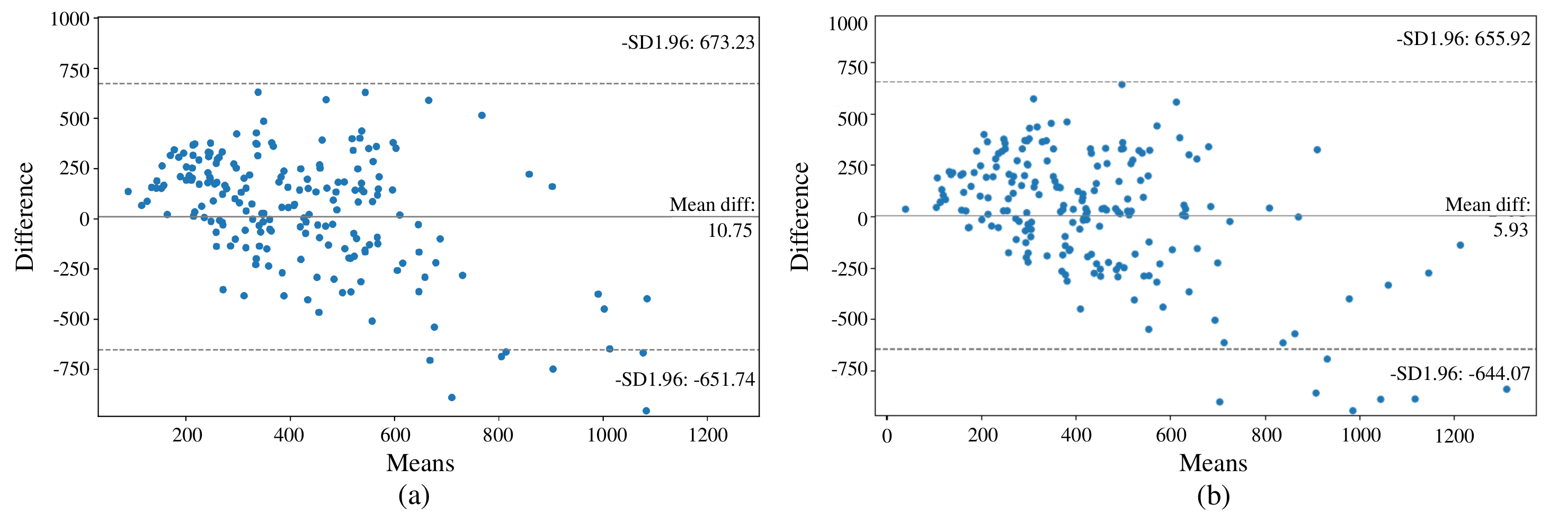}
  \caption{Bland Altman plot obtained from the training cross-validation results of overall survival prediction model  (a) Bland Altman plot obtained for all the extracted features. This gives a mean difference of 10.75 days.  (b) Bland Altman plot obtained for the selected 14 features. This gives a mean difference of 5.95 days. }
  \label{fig:mean_sd}  
\end{figure}

The results infer that our 3D attention UNet produces better accuracy than the original 3D UNet. Especially, the prediction of tumor core boosts up in our model (as shown in Fig. \ref{fig:result_comp}), which is a very important region to define tumor prognosis. To estimate the OS, we exploit 4 different regression models where XGBoost outperforms in terms of accuracy. To design an efficient model, we select the 14 most important features and train the models. A Bland Altman plot in Fig. \ref{fig:mean_sd} (a and b) represents the distribution of regression output for all extracted features and 14 selected features. The mean difference between the ground truth and the predicted survival rate is almost half (5.93 days) for the selected features comparing to all features. 

\begin{figure}[!h]
  \centering
  \includegraphics[width=0.9\linewidth]{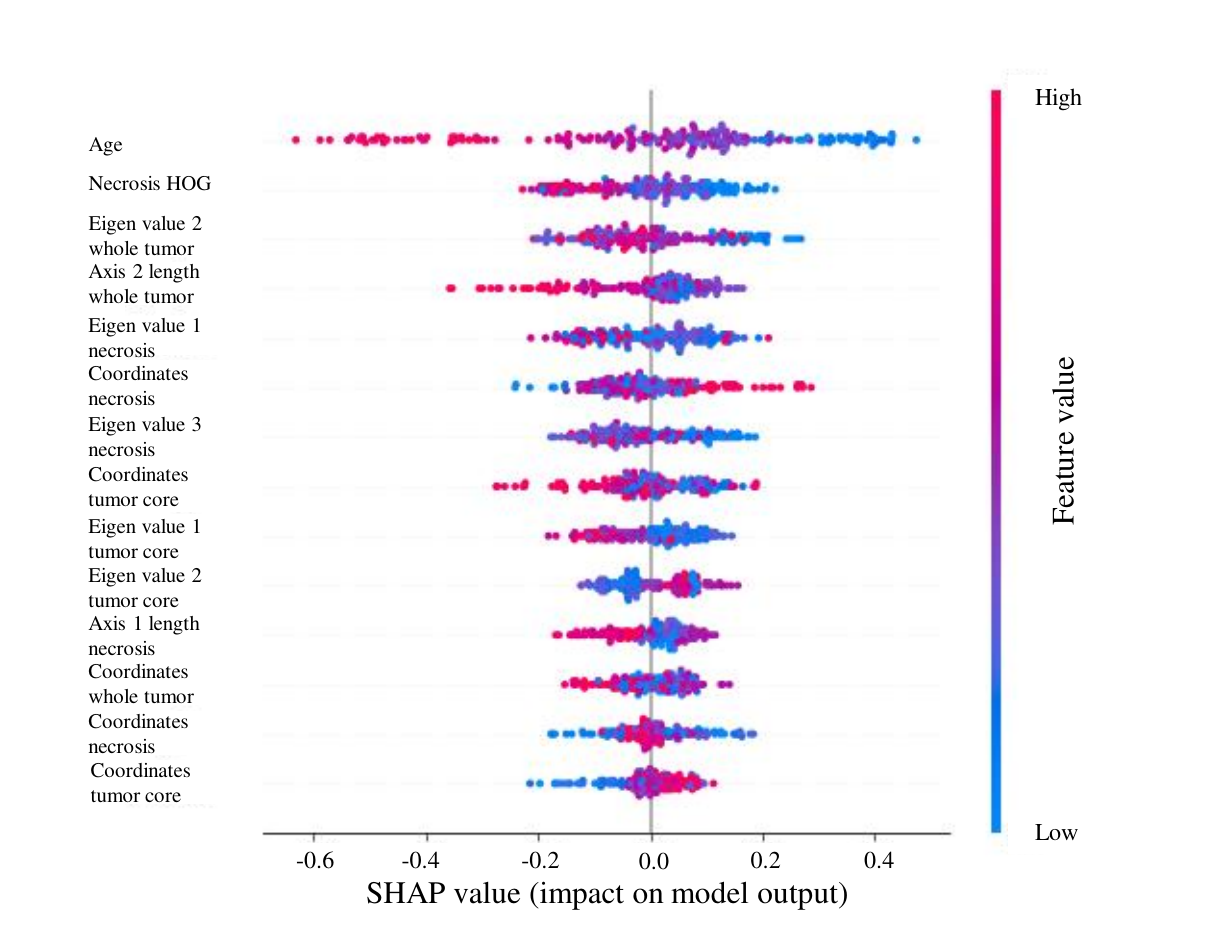}
  \caption{Effect of the features for the outcome of the model. The red color represents the high feature values, and blue represents the low values to determine the significance of the features in model prediction.}
  \label{fig:shap}  
\end{figure}

\begin{figure}[!htbp]
  \centering
  \includegraphics[width=0.5\linewidth]{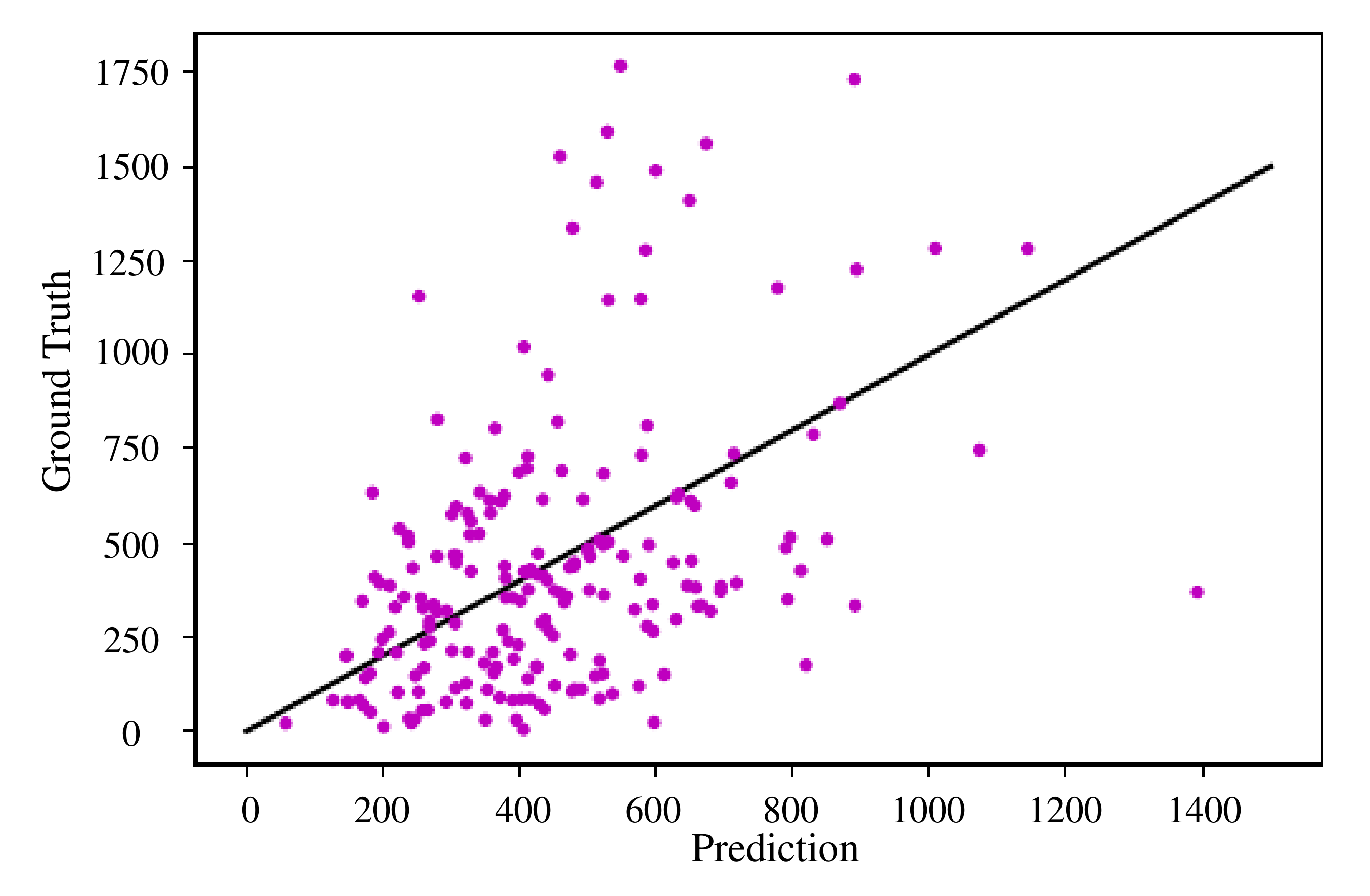}
  \caption{Regression scatter plot for the predicted overall survival and ground truth overall survival in days. }
  \label{fig:distribution}  
\end{figure}

Fig. \ref{fig:shap} demonstrate the importance of the selected features for the model performance. SHAP (SHapley Additive exPlanations) analysis, based on game theory, is an approach to explain the output of tree ensemble methods such as XGBoost. The red color represents the high feature values, and blue represents the low values. The y-axis of the plot shows the 14 features selected for our experiments. We can infer that age has the highest contribution to model performance. In addition, the histogram of necrosis, eigenvalue, whole tumor volume, and 2nd axis length of the tumor voxel are some of the significant features that contributed to predicting OS. Fig. \ref{fig:distribution} shows the regression plot of the ground truth and the prediction of the model.

\section{Conclusion}
In this paper, we present a segmentation and survival prediction model for automatic brain tumor prognosis using MRI. We adopt UNet and integrate the 3D attention technique into a novel way to capture the significant features in model learning. We also extract many novel geometric and shape features to estimate the survival days using the regression model. We observe that the location, shape, and size of the necrosis region is the most significant parameters in glioma prognosis estimation.

\section*{Acknowledgement}
This work was supported by the Singapore Academic Research Fund under Grant {R-397-000-297-114}, and NMRC Bedside \& Bench under grant R-397-000-245-511 awarded to Dr. Hongliang Ren.

%
%
\bibliography{mybib}{}

\begin{thebibliography}{10}
\providecommand{\url}[1]{\texttt{#1}}
\providecommand{\urlprefix}{URL }

\bibitem{bakas2017segmentation}
Bakas, S., Akbari, H., Sotiras, A., Bilello, M., Rozycki, M., Kirby, J.,
  Freymann, J., Farahani, K., Davatzikos, C.: Segmentation labels and radiomic
  features for the pre-operative scans of the tcga-gbm collection. The Cancer
  Imaging Archive  286 (2017)

\bibitem{bakas2017segmentationlgg}
Bakas, S., Akbari, H., Sotiras, A., Bilello, M., Rozycki, M., Kirby, J.,
  Freymann, J., Farahani, K., Davatzikos, C.: Segmentation labels and radiomic
  features for the pre-operative scans of the tcga-lgg collection. The Cancer
  Imaging Archive  (2017)

\bibitem{bakas2017advancing}
Bakas, S., Akbari, H., Sotiras, A., Bilello, M., Rozycki, M., Kirby, J.S.,
  Freymann, J.B., Farahani, K., Davatzikos, C.: Advancing the cancer genome
  atlas glioma mri collections with expert segmentation labels and radiomic
  features. Scientific data  4,  170117 (2017)

\bibitem{bakas2018identifying}
Bakas, S., Reyes, M., Jakab, A., Bauer, S., Rempfler, M., Crimi, A., Shinohara,
  R.T., Berger, C., Ha, S.M., Rozycki, M., et~al.: Identifying the best machine
  learning algorithms for brain tumor segmentation, progression assessment, and
  overall survival prediction in the brats challenge. arXiv preprint
  arXiv:1811.02629  (2018)

\bibitem{castells2009automated}
Castells, X., Garc{\'\i}a-G{\'o}mez, J.M., Navarro, A., Acebes, J.J., Godino,
  {\'O}., Boluda, S., Barcel{\'o}, A., Robles, M., Ari{\~n}o, J., Ar{\'u}s, C.:
  Automated brain tumor biopsy prediction using single-labeling cdna
  microarrays-based gene expression profiling. Diagnostic Molecular Pathology
  18(4),  206--218 (2009)

\bibitem{chen2016xgboost}
Chen, T., Guestrin, C.: Xgboost: A scalable tree boosting system. In:
  Proceedings of the 22nd acm sigkdd international conference on knowledge
  discovery and data mining. pp. 785--794. ACM (2016)

\bibitem{furnari2007malignant}
Furnari, F.B., Fenton, T., Bachoo, R.M., Mukasa, A., Stommel, J.M., Stegh, A.,
  Hahn, W.C., Ligon, K.L., Louis, D.N., Brennan, C., et~al.: Malignant
  astrocytic glioma: genetics, biology, and paths to treatment. Genes \&
  development  21(21),  2683--2710 (2007)

\bibitem{hu2018squeeze}
Hu, J., Shen, L., Sun, G.: Squeeze-and-excitation networks. In: Proceedings of
  the IEEE conference on computer vision and pattern recognition. pp.
  7132--7141 (2018)

\bibitem{islam2019real}
Islam, M., Atputharuban, D.A., Ramesh, R., Ren, H.: Real-time instrument
  segmentation in robotic surgery using auxiliary supervised deep adversarial
  learning. IEEE Robotics and Automation Letters  4(2),  2188--2195 (2019)

\bibitem{islam2018glioma}
Islam, M., Jose, V.J.M., Ren, H.: Glioma prognosis: Segmentation of the tumor
  and survival prediction using shape, geometric and clinical information. In:
  International MICCAI Brainlesion Workshop. pp. 142--153. Springer (2018)

\bibitem{islam2019learning}
Islam, M., Li, Y., Ren, H.: Learning where to look while tracking instruments
  in robot-assisted surgery. In: International Conference on Medical Image
  Computing and Computer-Assisted Intervention. pp. 412--420. Springer (2019)

\bibitem{islam2018ichnet}
Islam, M., Sanghani, P., See, A.A.Q., James, M.L., King, N.K.K., Ren, H.:
  Ichnet: Intracerebral hemorrhage (ich) segmentation using deep learning. In:
  International MICCAI Brainlesion Workshop. pp. 456--463. Springer (2018)

\bibitem{islam2018ischemic}
Islam, M., Vaidyanathan, N.R., Jose, V.J.M., Ren, H.: Ischemic stroke lesion
  segmentation using adversarial learning. In: International MICCAI Brainlesion
  Workshop. pp. 292--300. Springer (2018)

\bibitem{li2014medical}
Li, Q., Cai, W., Wang, X., Zhou, Y., Feng, D.D., Chen, M.: Medical image
  classification with convolutional neural network. In: 2014 13th International
  Conference on Control Automation Robotics \& Vision (ICARCV). pp. 844--848.
  IEEE (2014)

\bibitem{liaw2002classification}
Liaw, A., Wiener, M., et~al.: Classification and regression by randomforest. R
  news  2(3),  18--22 (2002)

\bibitem{louis20072007}
Louis, D.N., Ohgaki, H., Wiestler, O.D., Cavenee, W.K., Burger, P.C., Jouvet,
  A., Scheithauer, B.W., Kleihues, P.: The 2007 who classification of tumours
  of the central nervous system. Acta neuropathologica  114(2),  97--109 (2007)

\bibitem{menze2015multimodal}
Menze, B.H., Jakab, A., Bauer, S., Kalpathy-Cramer, J., Farahani, K., Kirby,
  J., Burren, Y., Porz, N., Slotboom, J., Wiest, R., et~al.: The multimodal
  brain tumor image segmentation benchmark (brats). IEEE transactions on
  medical imaging  34(10),  1993 (2015)

\bibitem{myronenko20183d}
Myronenko, A.: 3d mri brain tumor segmentation using autoencoder
  regularization. In: International MICCAI Brainlesion Workshop. pp. 311--320.
  Springer (2018)

\bibitem{paszke2017automatic}
Paszke, A., Gross, S., Chintala, S., Chanan, G., Yang, E., DeVito, Z., Lin, Z.,
  Desmaison, A., Antiga, L., Lerer, A.: Automatic differentiation in pytorch
  (2017)

\bibitem{ronneberger2015u}
Ronneberger, O., Fischer, P., Brox, T.: U-net: Convolutional networks for
  biomedical image segmentation. In: International Conference on Medical image
  computing and computer-assisted intervention. pp. 234--241. Springer (2015)

\bibitem{ruck1990multilayer}
Ruck, D.W., Rogers, S.K., Kabrisky, M., Oxley, M.E., Suter, B.W.: The
  multilayer perceptron as an approximation to a bayes optimal discriminant
  function. IEEE Transactions on Neural Networks  1(4),  296--298 (1990)

\bibitem{suykens1999least}
Suykens, J.A., Vandewalle, J.: Least squares support vector machine
  classifiers. Neural processing letters  9(3),  293--300 (1999)

\end{thebibliography}
\bibliographystyle{splncs03}

\clearpage
\end{document}